\documentclass[journal=nalefd,manuscript=article]{achemso}

\usepackage[T1]{fontenc}       
\usepackage{fancyhdr}

\newcommand{\onlinecite}[1]{\hspace{-1 ex} \nocite{#1}\citenum{#1}} 
\author{Johannes Binder}

\affiliation{Laboratoire National des Champs Magnetiques Intenses,
CNRS-UGA-UPS-INSA-EMFL, 25 Rue des Martyrs, 38042 Grenoble, France}
\alsoaffiliation{Faculty of Physics, University of Warsaw, Pasteura 5, 02-093 Warsaw, Poland}
\email{johannes.binder@fuw.edu.pl}

\author{Freddie Withers}
\affiliation
{School of Physics and Astronomy, University of Manchester, Oxford Road, Manchester M13 9PL, UK}
\alsoaffiliation{National Graphene Institute, University of Manchester, Oxford Road, Manchester, M13 9PL, UK}

\author{Maciej R. Molas}
\author{Clement Faugeras}
\author{Karol Nogajewski}
\affiliation{Laboratoire National des Champs Magnetiques Intenses,
CNRS-UGA-UPS-INSA-EMFL, 25 Rue des Martyrs, 38042 Grenoble, France}

\author{Kenji Watanabe}
\author{Takashi Taniguchi}
\affiliation{National Institute for Materials Science, 1-1 Namiki, Tsukuba 305-0044, Japan}

\author{Aleksey Kozikov}
\affiliation{School of Physics and Astronomy, University of Manchester, Oxford Road, Manchester M13 9PL, UK}
\alsoaffiliation{National Graphene Institute, University of Manchester, Oxford Road, Manchester, M13 9PL, UK}

\author{Andre K. Geim}
\affiliation{National Graphene Institute, University of Manchester, Oxford Road, Manchester, M13 9PL, UK}
\alsoaffiliation{Manchester Centre for Mesoscience and Nanotechnology, University of Manchester, Oxford Road, Manchester, M13 9PL, UK}

\author{Kostya S. Novoselov}
\affiliation{School of Physics and Astronomy, University of Manchester, Oxford Road, Manchester M13 9PL, UK}
\alsoaffiliation{National Graphene Institute, University of Manchester, Oxford Road, Manchester, M13 9PL, UK}

\author{Marek Potemski}
\affiliation{Laboratoire National des Champs Magnetiques Intenses,
CNRS-UGA-UPS-INSA-EMFL, 25 Rue des Martyrs, 38042 Grenoble, France}
\email{marek.potemski@lncmi.cnrs.fr}

\title[]{Sub-bandgap voltage electroluminescence and magneto-oscillations in a WSe$_2$ light-emitting van der Waals heterostructure}

\keywords{electroluminescence, magneto-oscillations, van der Waals heterostructures, tungsten diselenide, hexagonal boron nitride, acceptor}

\begin{document}




\begin{abstract}
We report on experimental investigations of an electrically driven WSe$_2$ based light-emitting van der Waals heterostructure. We observe a threshold voltage for electroluminescence significantly lower than the corresponding single particle band gap of monolayer WSe$_2$. This observation can be interpreted by considering the Coulomb interaction and a tunneling process involving excitons, well beyond the picture of independent charge carriers. An applied magnetic field reveals pronounced magneto-oscillations in the electroluminescence of the free exciton emission intensity with a 1/B-periodicity. This effect is ascribed to a modulation of the tunneling probability resulting from the Landau quantization in the graphene electrodes. A sharp feature in the differential conductance indicates that the Fermi level is pinned and allows for an estimation of the acceptor binding energy.

\end{abstract}


A new step of complexity has recently been undertaken in the field of two-dimensional crystals, by deterministically placing atomically thin layers of different materials on top of each other. The resulting stacks are referred to as van der Waals (vdW) heterostructures \cite{Geim2013, Novoselov2016}. Based on this idea, a few prototype devices, such as tunneling transistors \cite{Britnell2012,Georgiou2013,Zhao2015,Mishchenko2014,Lin2015,Roy2014,Greenaway2015,Gaskell2015, Wallbank2016} and/or light-emitting tunneling diodes \cite{Withers2015,Withers2015a,Clark2016,Palacios-Berraquero2016,Schwarz2016}, have been fabricated and successfully tested. However, further work is necessary in order to better characterize such structures, to learn more about their electronic and optical properties, with the aim to properly design device operation. 

Here, we unveil new facets of light emitting vdW heterostructures, with reference to the issue of the alignment of electronic bands, effects of Coulomb interaction and a subtle but still active role of the graphene electrodes in these devices. We report on optoelectronic measurements performed on a WSe$_2$-based tunneling light-emitting diode. The differential tunneling conductance of our structure shows a large zero bias anomaly (peak), which we ascribe to pinning of the Fermi energy at the WSe$_2$ impurity/acceptor level. A conceivable scenario for the evolution of the band alignment as a function of the bias voltage is proposed. Strikingly, the bias-potential onset for the electroluminescence is found to coincide with the energy of the free exciton of the WSe$_2$ monolayer (and not with the energy of a single-particle bandgap). This fact points out the relevant role of Coulomb interactions between electrically injected carriers on the tunneling processes in our device. Furthermore, pronounced magneto-oscillations are observed in the electroluminescence emission intensity measured as a function of magnetic field applied perpendicularly to the layer planes. These oscillations, periodic with the inverse of the magnetic field, reflect the modulation of the efficiency of carrier tunneling and are caused by the Landau quantization of the two-dimensional graphene electrodes.

\begin{figure}
\centering
\includegraphics{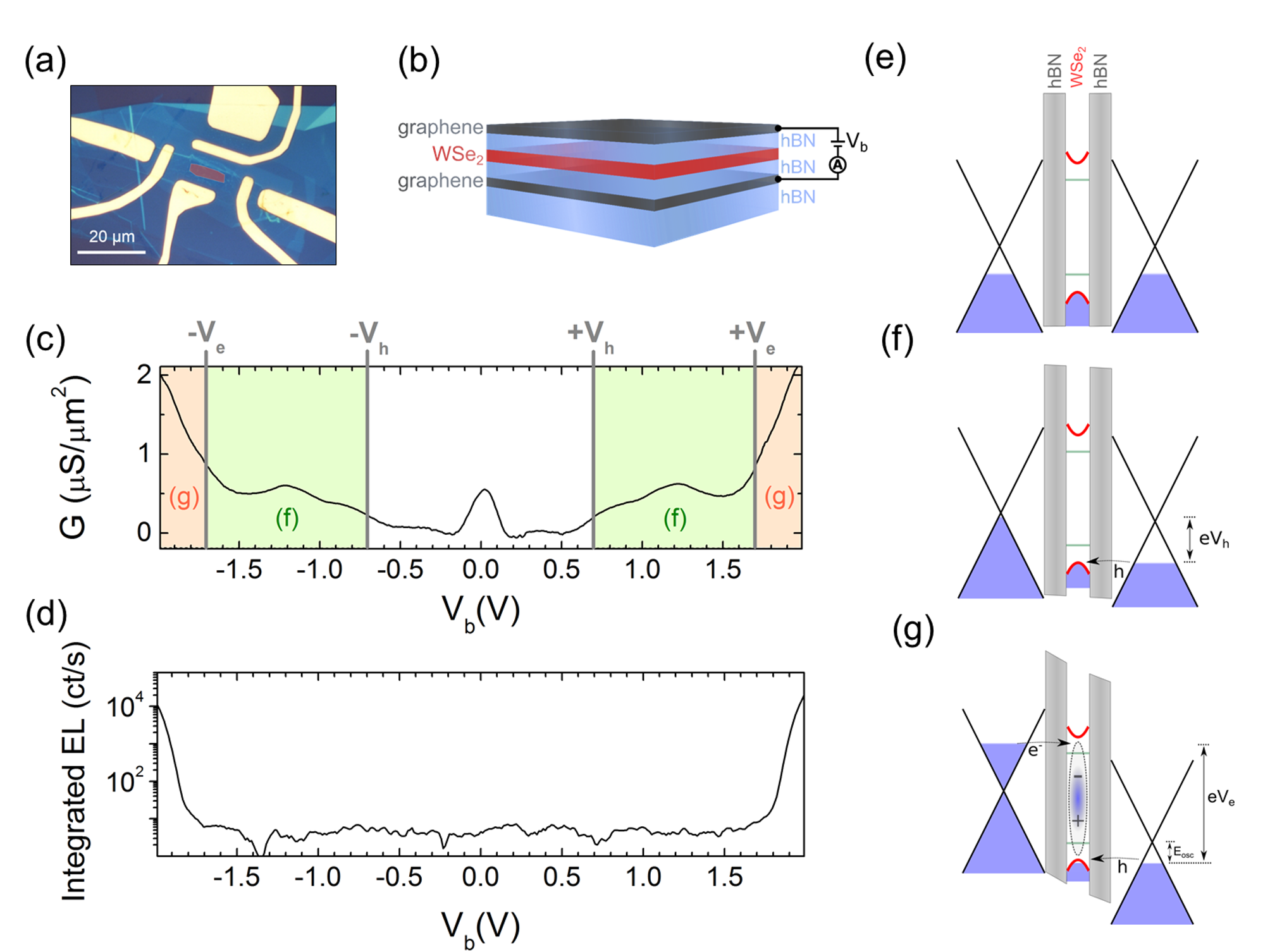}
\caption{Device structure and transport characteristics. (a)
Microscope image of the active part of the sample. The brown area
indicates the position of the WSe$_2$ flake. (b) Schematic
illustration of the heterostructure shown in (a). The layer
sequence from bottom to top is hBN/graphene/hBN/WSe$_2$/hBN/graphene. (c) Differential
conductance $G=dI/dV_b$. The colored regions correspond to the
situation shown in (f) and (g). (d) Integrated EL signal in a
spectral range from $1.48$~eV to $1.85$~eV as a function of the
bias voltage. (e) Schematic band diagram for zero bias. The red
parabolic bands correspond to the K-point of the Brillouin zone in
WSe$_2$. The horizontal green lines depict donor/acceptor-like
bands. (f) Schematic band diagram for the case of intermediate
bias. Here holes can tunnel through the hBN barrier into WSe$_2$,
but electrons can not. V$_h$ corresponds to the voltage required to
inject holes. (g) Band diagram for large bias. In this case, both
holes and electrons can tunnel. V$_e$ stands for the voltage
threshold for the tunneling of electrons into excitonic states of
WSe$_2$. The exciton is depicted schematically. In this regime light emission is observed due to
exciton recombination in the WSe$_2$ monolayer. E$_{osc}$
indicates the offset extracted from Eqn. \ref{eq:offset}.}
\label{fig:Fig1}
\end{figure}

We studied a light-emitting diode
structure~\cite{Withers2015,Withers2015a} that is based on a
WSe$_2$ monolayer as the active part. The layer sequence for this
device was Si / SiO$_2$ / hBN / graphene / hBN / WSe$_2$ / hBN / graphene. The emission
area of the structure is presented on the microscope image in
Figure \ref{fig:Fig1} (a). Figure
\ref{fig:Fig1} (b) depicts a schematic drawing of the
layered structure. The two hBN spacers that separate the WSe$_2$ monolayer
from the graphene electrodes are two layers thick. A detailed
description of the fabrication process can be found in Ref.~[\onlinecite{Withers2015}].

The optoelectronic characteristics of the sample were studied by recording the electroluminescence (EL) signal as a function of bias voltage and in magnetic fields up to 14 T. Current-voltage curves were measured to study the tunneling processes and photoluminescence (PL) mapping of the sample was performed to additionally characterize the structure. 
All measurements were performed with an optical-fiber-based insert placed in a superconducting coil. The investigated sample was located on top of an x-y-z piezo-stage kept in helium gas at $T=4.2$~K. The laser light from a continuous wave Ar+ laser ($\lambda=514.5$~nm) was coupled to an excitation fiber of $5~\mu$m diameter, focused on the sample by an aspheric lens. The signal was detected with a $50~\mu$m core fiber (collection spot diameter of $\sim 10~\mu$m) by a $0.5$~m long monochromator equipped with
a charge-couple-device (CCD) camera. Electrical measurements were performed using a Keithley 2400 source-measure unit.

A vertical current was observed upon the application of a bias voltage (V$_b$) between the two graphene electrodes. Such a charge transfer from one graphene layer to the other can only be achieved via tunneling. To describe the electronic transport
perpendicular to the structure, we present in
Figure~\ref{fig:Fig1} (c) the variations of the
differential conductance G=dI/dV$_b$ as a function of the bias
applied between the two graphene electrodes. The optical emission was monitored at the same time. The corresponding integrated EL intensity is presented in Figure~\ref{fig:Fig1} (d).

To interpret the optoelectronic behavior of this device,
it is crucial to know the band alignment of the heterostructure. To this end one has to rely both on the theoretical
estimations~\cite{Gong2013,Kang2013} and on the spectroscopic
experimental works targeting these band
offsets~\cite{Britnell2012,Zhao2015,Wilson2016}. A schematic illustration of the bands
is shown in Figure \ref{fig:Fig1} (e). The drawing
depicts the two graphene electrodes, represented by Dirac cones,
which are separated from the WSe$_2$ monolayer by the hBN
barriers. The WSe$_2$ layer is schematically illustrated by the
parabolic bands around the K-point of the Brillouin zone. In
addition donor/acceptor-like bands are depicted by horizontal
green lines in WSe$_2$. Transport measurements of the tunneling
current in hBN/graphene/hBN structures found the valence band of
hBN to be offset by about $1.4-1.5$~eV
\cite{Britnell2012,Zhao2015} from the graphene Dirac point. The
band alignment of monolayer WSe$_2$ and graphene has been recently
studied using $\mu$-ARPES and an offset of $0.70$~eV between
the Dirac point and the WSe$_2$ valence band edge has been reported~\cite{Wilson2016}. Assuming a direct band gap for monolayer
WSe$_2$ of about $2-2.2$~eV \cite{He2014,Zhang2015} one can
conclude that the energy separation between the Dirac point of
graphene and the valence band edge of WSe$_2$ should be
significantly lower as compared to the conduction band. This
finding is also in agreement with theoretical estimations
\cite{Gong2013,Kang2013}. The results are summarized qualitatively
in the sketch in Figure \ref{fig:Fig1} (e).

By using the proposed band alignment scenario, we can divide the differential conductance (Figure~\ref{fig:Fig1} (c)) in three distinct tunneling regimes. The first one occurs at around zero bias, where a pronounced peak is observed.
We ascribe this feature to be caused by the tunneling
through impurity donor/acceptor bands in WSe$_2$, that pin the
Fermi level (Figure~\ref{fig:Fig1} (e)). With an
increase in bias voltage the tunneling through the impurity band
ceases to be resonant and a decrease in differential conductance
is observed, giving rise to a symmetric peak-like shape. Our measurements cannot directly infer whether these impurities are
of donor or acceptor type. However, within the expected band
alignment, the Dirac point of graphene is much closer in energy to
the valence band edge of WSe$_2$, a material which shows preferably
p-type conductivity
\cite{Fang2012,Pradhan2015,Movva2015,Fallahazad2016,Campbell2016}.
Hence, we assume that the dominant impurities in the investigated WSe$_2$ monolayer are
of acceptor type, and that the Fermi level is pinned to this
impurity band at zero applied bias. A small applied bias is then sufficient to move the
Fermi levels of the graphene electrodes out of resonance with this
band, producing a symmetric differential conductance feature
centered at zero applied bias.
The peak at zero bias was also observed for other similar WSe$_2$ devices, however it was found that its magnitude can strongly vary from device to device (see supporting information) and it can also be absent. This variation can be understood in terms of different unintentional initial doping of WSe$_2$, which might vary from flake to flake.

The second regime, indicated by the green color in
Figure~\ref{fig:Fig1} (c), shows an overall increase of
conductance with two peak-like features. The first feature, around
$V_h \sim \pm 0.7$~V, originates from the onset of hole
tunneling into the valence band of WSe$_2$. This process becomes
efficient when the Fermi level of one graphene layer is moved by
the amount of the acceptor binding energy, to coincide with the
valence band edge of WSe$_2$. The situation is schematically depicted in
Figure~\ref{fig:Fig1} (f). At this point one should mention that another possible reason for the above mentioned peak at zero bias could be resonant effects due to the direct graphene-graphene tunneling \cite{Britnell2013}. However, the graphene electrodes were not intentionally aligned, making the appearance of resonant effects very improbable. Another argument against this alternative scenario is that with a Femi level close to the Dirac point, one would roughly need to apply a voltage corresponding to twice the valence band offset to enable hole tunneling, which does not fit the observation of V$_h \sim \pm 0.7$~V. The second feature in this
regime is the peak at larger bias voltage ($V_d \sim \pm 1.2$~V).
The origin of this peak is still unclear, and a possible explanation could be tunneling
involving mid-gap impurity states in WSe$_2$.

The above discussion yields three conditions: an onset voltage for hole tunneling of $V_h \sim \pm 0.7$~V, a
valence band offset for monolayer WSe$_2$ of E$_{VB}\sim 0.7$~eV, and a Fermi level that is pinned at zero applied bias to the acceptor level. These conditions together with simple considerations regarding the band structure and the electric field in the sample allows us to estimate an acceptor binding energy of $E_{acc} \sim 250$~meV (see supporting information).

At larger bias (third regime) the increase in voltage will mostly drop across the
graphene / hBN junction that does not permit tunneling into the
WSe$_2$ layer. In order to observe EL, both electrons and holes
must be present in the WSe$_2$ layer. This condition is satisfied
in the voltage region around $V_e \sim \pm 1.7$~V above which EL is observed. The voltage dependence of the spectrally integrated EL
signal, shown in Figure \ref{fig:Fig1} (d), displays
a steep onset of emission in that bias range. We can therefore
ascribe the strong increase in conductance to the tunneling of
electrons into the WSe$_2$ monolayer (compare Figure
\ref{fig:Fig1} (g)). Additional data for a similar device showing the same behavior is presented in the supporting information.
Strikingly, the onset for EL of
$V_e \sim \pm 1.7$~V is significantly smaller as compared to the
direct band gap of a WSe$_2$ monolayer which is of about
$2-2.2$~eV \cite{He2014,Zhang2015}. Because the base temperature
of our experiment $T=4.2$~K implies a thermal energy below
$400$~$\mu$eV and given the relative alignment of the graphene
electronic bands with respect to those of hBN, the large difference
can hardly be explained in terms of thermal activation of carriers
or a lowering of the effective hBN barrier caused be the electric
field. However, the EL onset at about $V_e \sim \pm 1.7$~V
corresponds well with the emitted free exciton energy of $\sim 1.72$~eV. Based on our experiments, the most probable scenario involves tunneling\cite{Geim1994} directly
into the excitonic states of the WSe$_2$ monolayer. Because the tunneling of
holes starts at bias voltages close to $V_h \sim \pm 0.7$~V,  a
population of holes is already present in the valence band when
electrons start to tunnel, directly forming excitons. Such
processes were indeed observed for resonant electron tunneling
into p-doped GaAs quantum wells (QWs) \cite{Cao1995}. In the case of WSe$_2$ monolayer, the
exciton binding energies are large ($\sim 0.4$~eV \cite{He2014})
as compared to excitons in GaAs QW systems, which gives rise to
the observed large differences. Moreover, it was shown that excitons can persist in such materials up to large carrier concentrations \cite{Scharf2016}, with an estimation of several
$10^{13}$~cm$^{-2}$ required for the quenching of the excitonic resonances \cite{Chernikov2015}.

\begin{figure}
    \centering
        \includegraphics{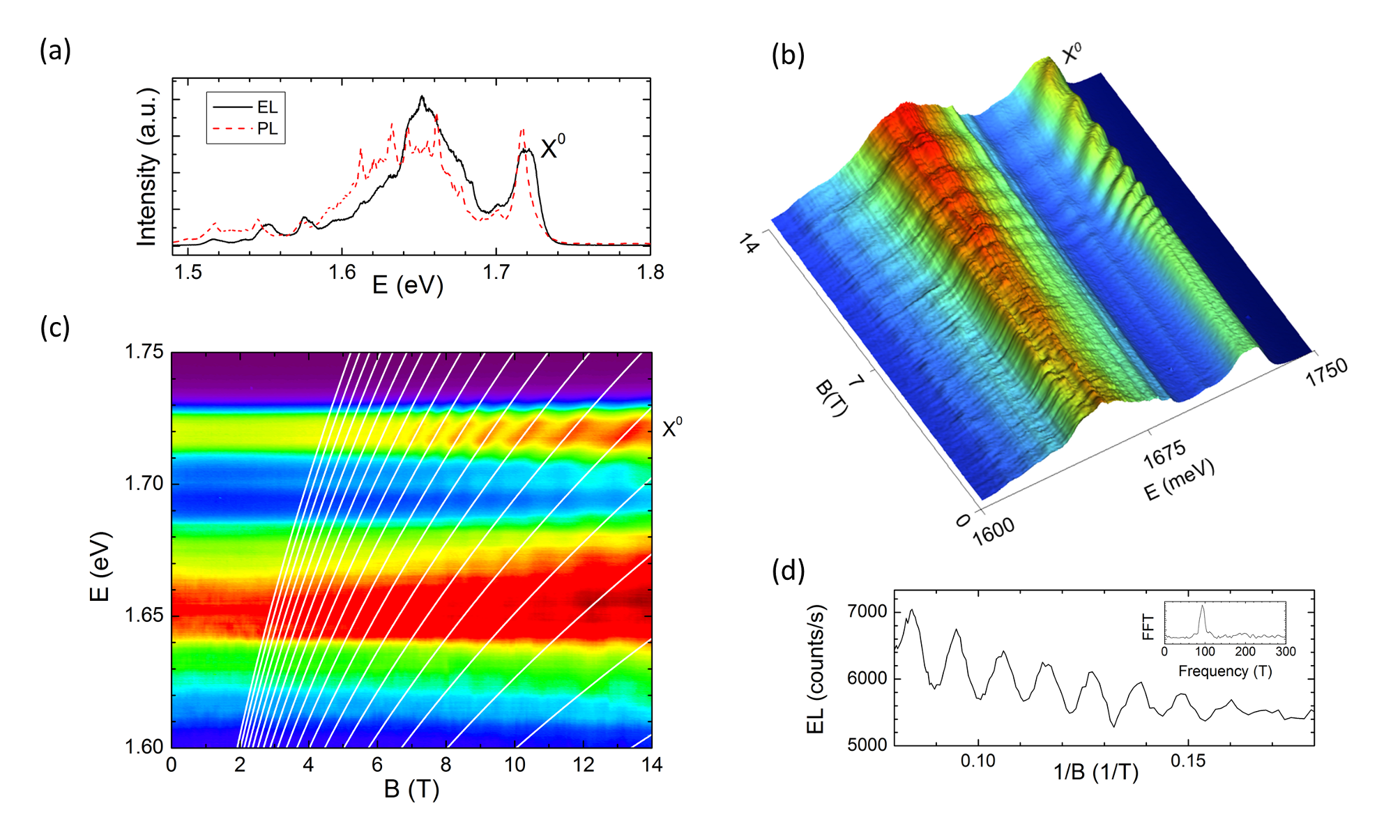}
    \caption{Optical characteristics in magnetic field. (a) Representative
EL - (black line) and PL - (red dashed line) spectra for $B = 0$.
(b) Three dimensional false color plot of the raw EL signal as a
function of the magnetic field for a constant current of
I=$36$~$\mu A$. (c) Overlay of the two-dimensional EL false color
map with the graphene Landau level spectrum (white lines). The
graphene LL spectrum was calculated using $v_{f}=1 \cdot 10^6
\frac{m}{s}$ and an energy offset for the Dirac point of
$E_{osc}=350$~meV as deduced from the observed periodicity of
$93$~T. (d) Horizontal cut to (c) at $E=1.72$~eV (neutral exciton)
as a function of $1/B$. Inset: result of the Fourier analysis that
shows a sharp peak at $93$~T.}
    \label{fig:Fig2}
\end{figure}

Figure \ref{fig:Fig2} (a) shows representative
PL and EL spectra. The highest energy band ($E \sim 1.72$~eV) labelled
X$^0$ can be attributed to the neutral, free A exciton resonance.
As observed in EL, the X$^0$ feature has a full width at half
maximum (FWHM) close to $20$~meV, hence $3$ to $4$ times bigger than in PL (red dashed line in Fig. \ref{fig:Fig2} (a)). The large FWHM originates from inhomogeneous broadening, which is more apparent in EL than PL since in the case of the former the signal is collected from the entire flake. At lower energies, a complex broad band is observed, typical for
monolayer WSe$_2$ samples and which has been attributed to charged
and localized (bound) excitons~\cite{Jones2013,Withers2015a,Arora2015,Koperski2015}. This large broad band indicates the presence of a significant amount of impurities in the case of our device. The presence of defects as evidenced by the optical measurements fits into the scenario of a pinned Fermi level for this device. A
magnetic field was applied perpendicular to the surface
of the structure in order to study its impact on the EL signal.
First, a strong magneto-resistance develops in the structure and
significantly shifts the threshold bias for EL emission to
larger voltages with increasing applied magnetic field. We ascribe this additional resistance to be
caused by the in-plane magneto-resistance of the graphene
contacts, which serve as conductors between the metal contacts and
the active area~\cite{Greenaway2015}. This effect hinders
measurements with constant applied voltage. To compensate for the additional voltage drop a constant
current was kept for the magnetic field sweeps, which yielded
stable EL measurement conditions.

Figure \ref{fig:Fig2} (b) shows a three-dimensional false
color plot of the raw EL signal as a function of magnetic field.
We observed a very intense modulation of the X$^0$ line with
an intensity and a shape that varies as a function of the magnetic field. 
This modulation of the exciton emission is not a simple on-off effect, but due to the large width of the
$X^0$ feature in EL, an energy-dependent
modulation of the X$^0$ emission can be observed. In order to establish the origin
of the modulations, a cut at a constant energy of the
X$^0$-feature plotted as $1/B$ is presented in Figure
\ref{fig:Fig2} (d). A $1/B$-periodicity is apparent, which
is further supported by the results of a Fourier analysis for this
graph giving a well defined peak for a period $\Delta(1/B)=93$~T (see
inset). Assuming Landau quantization in the graphene
electrodes to be responsible for the observed behavior, one
obtains the following Landau level (LL) spectrum
\cite{Faugeras2011,Faugeras2014}

\begin{equation}
    E_n=sign(n)v_f \sqrt{2e\hbar B |n| }
    \label{eq:LL}
\end{equation}

where $v_f=1 \cdot 10^6 \frac{m}{s}$ is the Fermi velocity and $n$
the Landau level index. For a constant energy cut across the
LL-spectrum of graphene, one obtains oscillations with a $1/B$
periodicity. Hence, by using the extracted periodicity one can
determine the energy above the Dirac point by calculating

\begin{equation}
    E_{osc}=v_f\sqrt{\frac{2e \hbar}{\Delta(\frac{1}{B})}}
    \label{eq:offset}
\end{equation}

\begin{figure}
    \centering
        \includegraphics{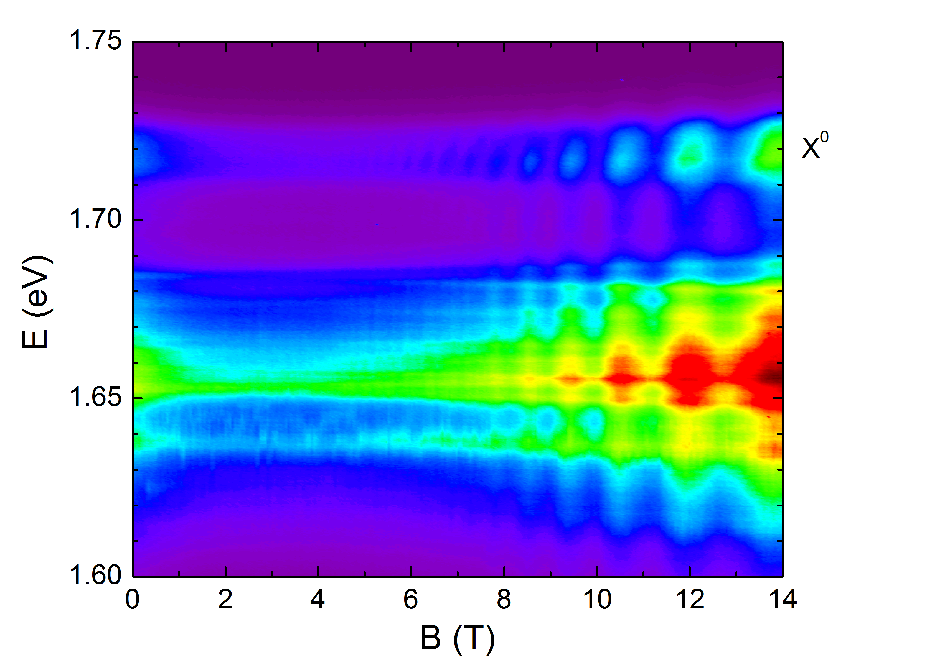}
    \caption{Oscillations of the localized excitonic bands. Two-dimensional
false color map of the magnetic field dependence of the EL signal
for a constant current of $I=30$~$\mu A$. The vertical streaks
indicate the transfer of periodicity from the free exciton line
X$^0$ to the localized excitonic bands. }
    \label{fig:Fig3}
\end{figure}

This consideration yields an energy separation of about
E$_{osc}=350$~meV. In Figure \ref{fig:Fig2} (c) we present
an overlay of the graphene Landau levels with the Dirac point
located $350$~meV below the energy of the X$^0$ line and the
measured EL-spectra. We find an excellent agreement, since the
spacing as well as the energy dependence of the modulations are
fully described. Consequently, we conclude that the oscillations
are related to the quantized density of states (DOS) of the graphene electrodes. This
quantization leads to oscillations in the population of holes in
the WSe$_2$ valence band, since the injection process via
tunneling from graphene is modulated by the LL spectrum.
At lower energies, no signatures of an energy-dependent modulation
could be observed for the broad localized (bound) exciton emission
band. However, an attenuation of the broad band that oscillates with the magnetic field but not with the energy is apparent
when the exciton is affected
by the above discussed tunneling process. Such a behavior can be
observed in Figure \ref{fig:Fig2} (b) and (c) in the form
of lines on top of the broad emission band. This effect is more
markedly shown in Figure \ref{fig:Fig3}, which
presents an EL false color map at lower injection current. 

This effect confirms that the population of these localized
states is fed by the population of free excitons: electrons are
injected into the WSe$_2$ layer via tunneling and directly bound
to holes to first form excitons which can scatter to the localized
excitonic bands, at lower energy~\cite{Wang2014}. As a
consequence, instead of an energy-dependent modulation, as it was
observed for the broadened X$^0$ line, one would expect the
oscillation frequency to be transferred from the free exciton to
the localized bands.

Another characteristic observation is the appearance of a single frequency
although there are two tunneling processes (holes and electrons)
with different tunneling barriers. Eqn.~\ref{eq:LL} describes a
Landau level fan chart with an energy spacing between consecutive
LLs that decreases with increasing LL-index n. This implies that
for a high Fermi energy, the LL spectrum can be significantly smeared out, thus no clear oscillatory behavior can be
expected concerning the electron tunneling. This observation of a
single frequency modulating the exciton emission intensity, can
therefore be seen as an additional evidence that the graphene
electrodes are responsible for the oscillations.
The modulation in tunneling that was registered optically should in
principle also be present in the electrical characteristics.
Instead, the experiment shows no clear modulation in the
measured voltage as a function of magnetic field. This discrepancy
can be understood when taking into account the actual processes
that influence the two measurement techniques. The EL signal is
only sensitive to the excitonic population, which is a result of
the injection of electrons and holes via tunneling. The
electrical measurement, however, is a sum of all possible
tunneling pathways and does also include leakage and parasitic
components, which can mask the effect. With the magneto-EL
measurement we therefore gained information difficult to access
with standard magneto-transport tunneling experiments.

In summary, we report on optoelectronic properties of a WSe$_2$ based tunneling light-emitting vdW heterostructure in magnetic fields. We propose a conceivable scenario for the band alignment in the structure, which allows us to estimate an acceptor binding energy. 
The Landau quantization in the graphene electrodes is shown to strongly
modulate the injection of holes into the valence band of the
active WSe$_2$ monolayer, which in turn modulates the EL signal.
The observed oscillations of the neutral exciton intensity as a
function of the magnetic field show a pronounced 1/B periodicity
which was used to deduce an effective band offset between
graphene's Dirac point and the valence band edge of the WSe$_2$
monolayer. Our results hence show that the role of graphene electrodes in vdW
heterostructures goes far beyond being a semitransparent electrode with a low density of
states.\\
In addition, we observed EL emission for applied voltages well below
the corresponding band gap of monolayer WSe$_2$, which was explained in
terms of direct tunneling of carriers into excitonic states in
WSe$_2$. We found the EL signal to be more
sensitive to the quantized hole injection as compared to
magneto-transport, which illustrates the advantage of optoelectronic
tunneling measurements.
Our findings highlight the importance of
excitonic states for the tunneling processes in vdW heterostructures, 
giving rise to sub-bandgap EL, which could be a key aspect for future optoelectronic device engineering.

\begin{acknowledgement}

This work was supported by European Research Council Synergy Grant Hetero2D, EC-FET European Graphene Flagship (no.604391), The Royal Society, Royal Academy of Engineering, U.S. Army, Engineering and Physical Sciences Research Council (UK), U.S. Office of Naval Research,  U.S. Air Force Office of Scientific Research and the European Research Council (MOMB project no.320590).

\end{acknowledgement}

\section{Supporting Information}
Band structure and electric field considerations, additional data on the zero bias anomaly on other samples, evolution of the electroluminescence as a function of bias voltage for a different sample.

\clearpage

\section{Supporting information}

\subsection{Band structure and electric field considerations}

This section describes the approach employed to obtain an estimation for the acceptor binding energy given in the main text (E$_{acc}\sim 250$~meV). 
To extract this value we made the following assumptions:

\begin{itemize}
    \item The electric field is homogeneous across the structure for voltages below and equal to $V_h$, i.e. no free carriers are present in the WSe$_2$ monolayer. This condition makes it reasonable to simplify the situation by introducing an effective dielectric constant weighted by the thicknesses of the layers to describe the electric fields.
        
    \item The offset between the Dirac point of the graphene electrodes and the valence band of the WSe$_2$ layer is $E_{VB}=-0.7$~eV. As described in the main text this value is based on literature.
        
        \item The Fermi level is pinned to the acceptor states at zero applied voltage, in accordance with our interpretation of the feature centered at $0$~V observed in the differential conductance (see main text).
        
    \item The onset of hole tunnelling corresponds to a voltage of $V_h= \pm 0.7$~V, as extracted from our measurements.
\end{itemize}
 
As a first step we define an effective dielectric constant $\epsilon_{eff}$ for the whole hBN/WSe$_2$/hBN stack (using $\epsilon_{hBN}=4$\cite{Kim2012,Young2012}, $\epsilon_{WSe_2}=7.2$ \cite{Kim2015,Ghosh2013},$d_{hBN}=1.34$~nm, $d_{WSe_2}=0.65$~nm)

\begin{equation}
    \epsilon_{eff}=\frac{\epsilon_{hBN} \cdot d_{hBN}+\epsilon_{WSe_2}\cdot d_{WSe_2}}{d_{stack}} \sim 5
    \label{eq:effective_dielectric}
\end{equation}

Where $\epsilon_{hBN}$, $\epsilon_{WSe_2}$ are the dielect contants of hBN and of WSe$_2$, respectively. We use $\epsilon_{eff}$ to estimate the electric field F according to

\begin{equation}
    F=\frac{V_h}{d_{stack} \cdot \epsilon_{eff}}.
\end{equation}

This field causes an energy shift of the Dirac cones, which can be written as

\begin{equation}
    E_{field}=e \cdot F \cdot d_{stack}=\frac{e \cdot V_h}{\epsilon_{eff}}
\end{equation}

\begin{figure}
    \centering
        \includegraphics{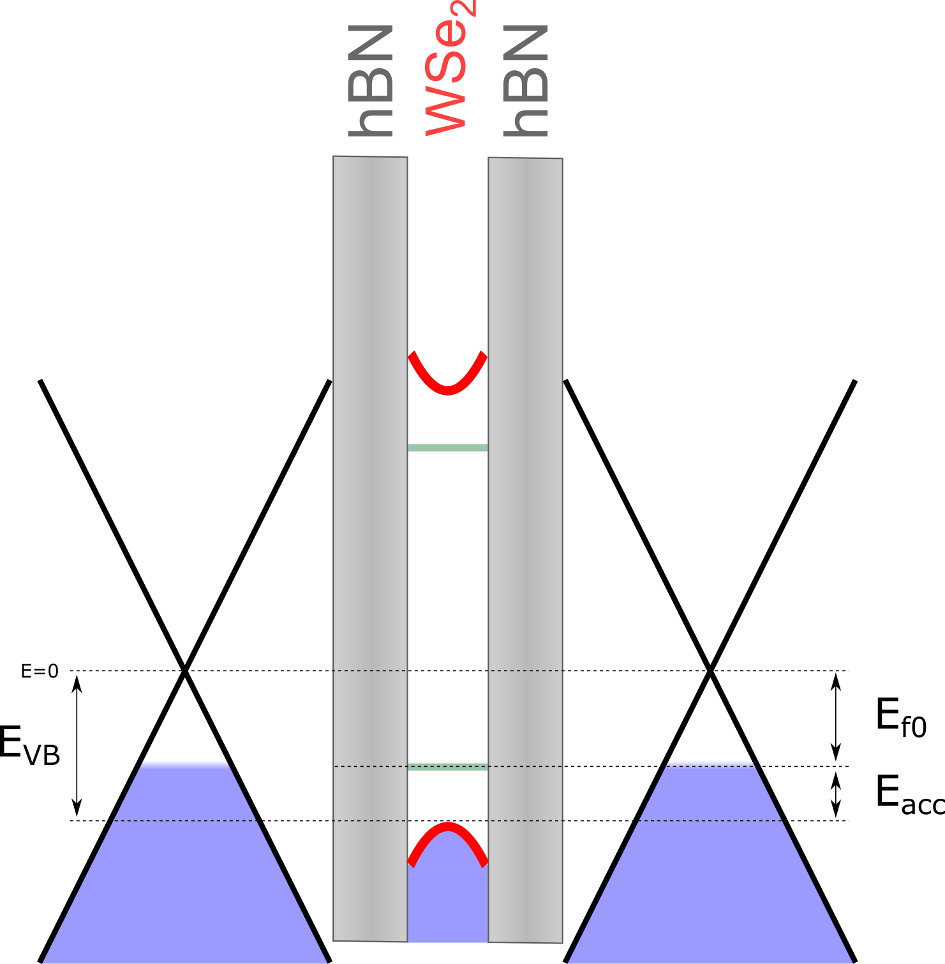}
    \caption{Definition of the relevant energies.}
    \label{fig:FigSI1}
\end{figure}

The definitions of the relevant energies needed to estimate the
acceptor energy $E_{acc}$ are presented in Figure \ref{fig:FigSI1}.

\begin{figure}
    \centering
        \includegraphics{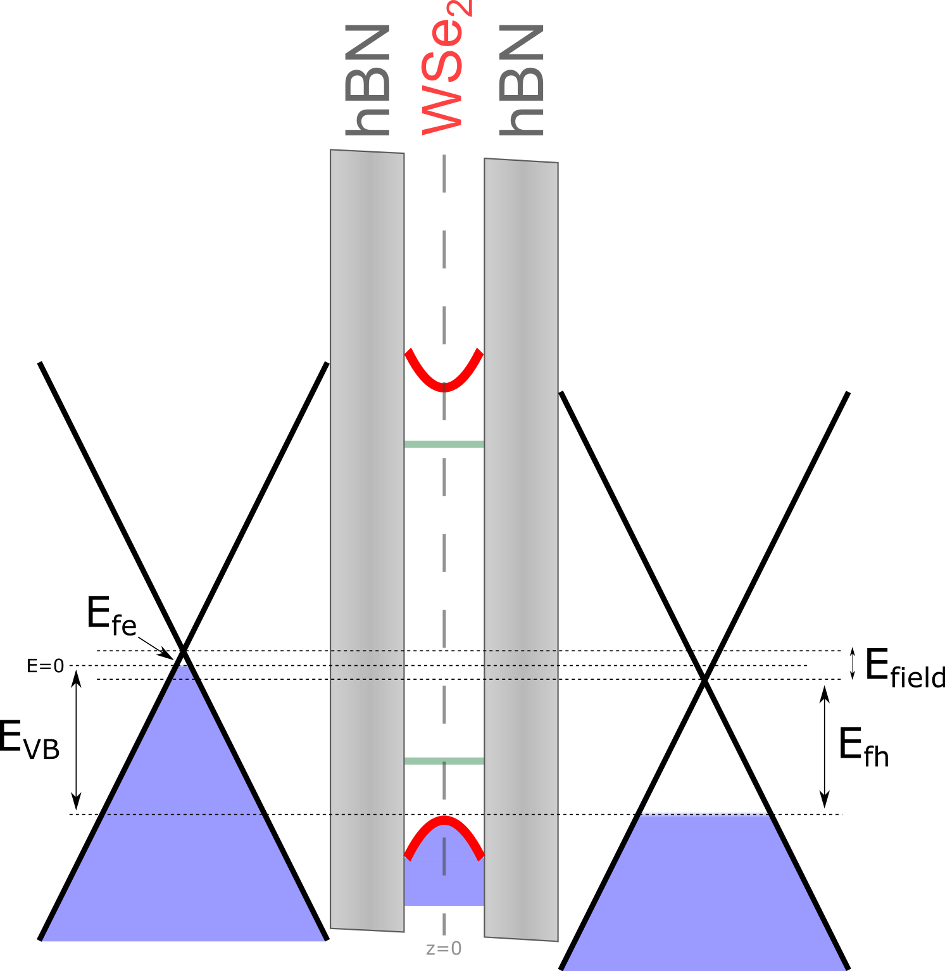}
    \caption{Schematic band structure for an applied voltage of $V_h=0.7$~V.
    The dashed vertical line displays the symmetry plane and indicates the offspring.}
    \label{fig:FigSI2}
\end{figure}

The application of a voltage equal to $V_h=\pm 0.7$~V leads to the build-up of an electric field across the structure (see Figure \ref{fig:FigSI2}). As assumed above we estimate $E_{field}$ by dividing the applied voltage by the effective dielectric constant giving $E_{field}=0.14$~eV.

The Dirac cones of the graphene electrodes should shift by the same value equal to $|E_{field}/2|$, since we assumed the screening inside the stack to be zero. Please note the we choose the origin to be exactly in the center of the symmetric structure (long dashed line in Figure \ref{fig:FigSI2}). We now can calculate the Fermi energy for holes $E_{fh}$, since we know that it has to coincide with the valence band edge of WSe$_2$ for the applied voltage $V_h$.

\begin{equation}
    E_{fh}=|E_{VB}|-|\frac{E_{field}}{2}|=-0.63~\mbox{eV}
\end{equation}

which gives a hole concentration ($v_f=1 \cdot 10^6$~m/s) of

\begin{equation}
    n_h=2.92 \cdot 10^{13}~\frac{1}{\mbox{cm}^2}
\end{equation}

The Fermi level in the second graphene electrode can be obtained since we apply a constant voltage, which is equal to a constant energy difference between the quasi Fermi levels $E_{fh}$ and $E_{fe}$. By using this fact we obtain

\begin{equation}
    E_{fe}=|e\cdot V_h|-|E_{field}|-|E_{fh}|= -0.07~\mbox{eV}
\end{equation}

and hence we obtain a hole concentration of

\begin{equation}
    n_e=3.54 \cdot 10^{11}~\mbox{cm}^{-2}.
\end{equation}

The charge conservation allows us to obtain the initial hole carrier concentration of the pinned Fermi level, which yields

\begin{equation}
    n_0=\frac{n_e+n_h}{2}=1.48 \cdot 10^{13}~\mbox{cm}^{-2}.
\end{equation}

This concentration corresponds to an initial Fermi level $E_{f0}$ of

\begin{equation}
    E_{f0}=-0.45~\mbox{eV}.
\end{equation}

Having obtained the position of the pinned Fermi level we can estimate the acceptor binding energy to be

\begin{equation}
    E_{acc}=|E_{VB}|-|E_{f0}|=0.25~\mbox{eV}.
\end{equation}

\newpage

\subsection{Zero bias anomaly}

\begin{figure}[H]
	\centering
		\includegraphics[width=0.60\textwidth]{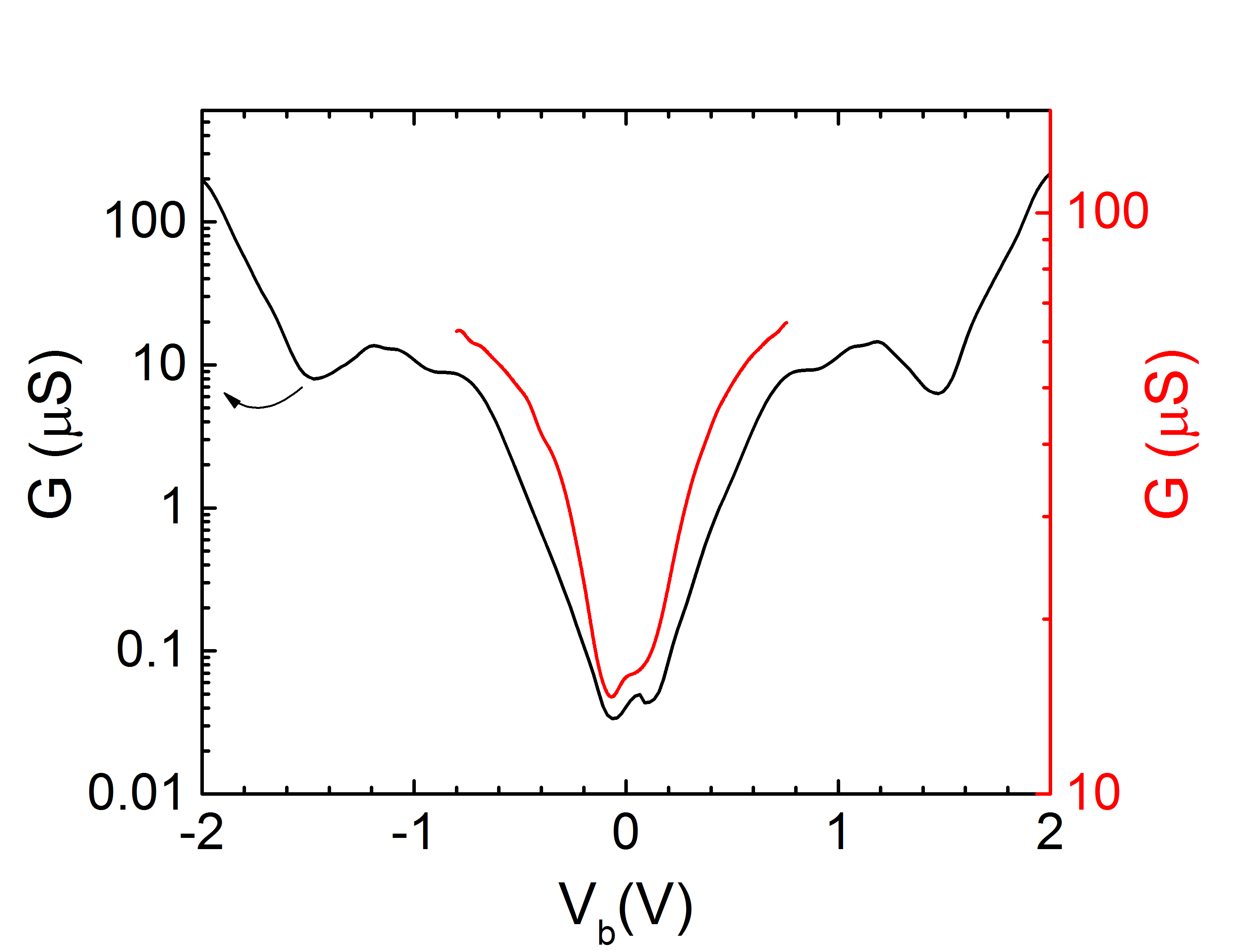}
	\caption{Differential conductance curves of two additional devices, showing a zero bias anomaly.}
	\label{fig:FigSI3}
\end{figure}

As mentioned in the main text a peak at zero bias was also observed for other similar graphene/hBN/WSe2/hBN/graphene heterostructures. Figure \ref{fig:FigSI3} presents the differential conductance curves for two additional devices showing this peak.

\subsection{Additional electroluminescence data}

\begin{figure}[h]
	\centering
		\includegraphics[width=0.70\textwidth]{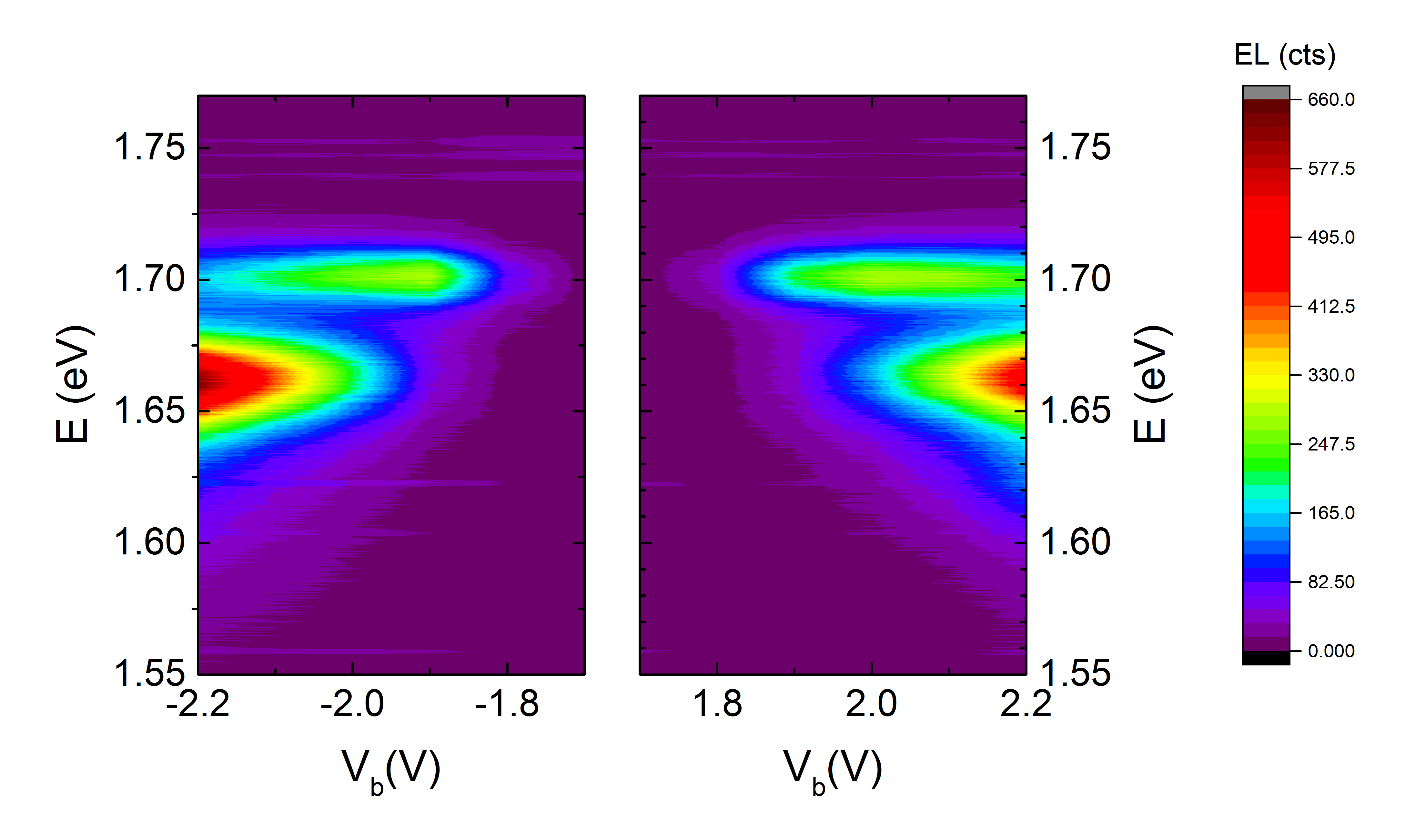}
	\caption{False color map of the electroluminescence versus bias voltage behavior for another device. The electrical charateristics of this sample are presented as black trace in Figure \ref{fig:FigSI3} ($T=10$~K).}
	\label{fig:FigSI4}
\end{figure}

Figure \ref{fig:FigSI4} presents the electroluminescence (EL) versus bias voltage behavior for another device. The differential conductance curve for this device is presented above as black trace in Figure \ref{fig:FigSI3}. The map in Figure \ref{fig:FigSI4} clearly illustrates that the onset for EL is slightly above $V\sim\pm 1.7$~V in agreement with the interpretation of excitonic states taking part in the tunneling as given in the main text.

\begin{figure}[H]
	\centering
		\includegraphics[width=0.70\textwidth]{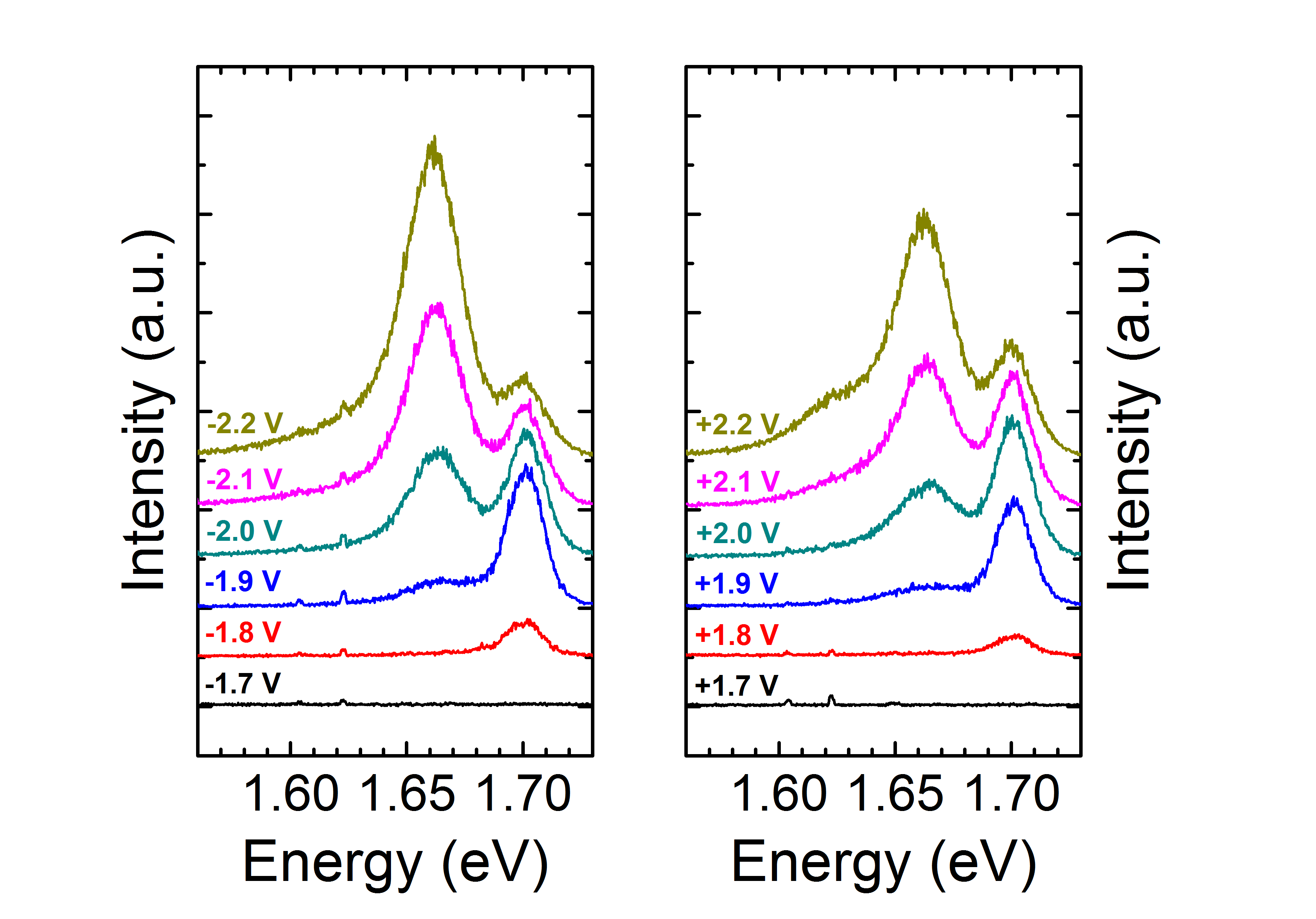}
	\caption{Electroluminescence spectra for different bias voltages corresponding to Figure \ref{fig:FigSI4}. The spectra were shifted vertically for clarity.}
	\label{fig:FigSI5}
\end{figure}

Figure \ref{fig:FigSI5} shows the EL spectra for several bias voltages corresponding to the map in Figure \ref{fig:FigSI4}. For $V\sim\pm 1.8$~V a clear EL signal can be observed showing that the onset is well below the band gap of monolayer WSe$_2$ of about $2-2.2$~eV \cite{He2014,Zhang2015}.

The bulk crystal used for the exfoliation of monolayer WSe$_{2}$ was purchased from HQ graphene.

\bibliography{references}

\end{document}